# Peran Pemerintah Pusat dan Daerah dalam Menyediakan Pendidikan Dasar Bermutu untuk Mewujudkan Visi Indonesia 2045

Ngan Sui-Ni [1(a)]

[1] Dinas Pendidikan dan Kebudayaan Kabupaten Bangka Selatan, Indonesia

[a)] ani_dikbud@bangkaselatankab.go.id

| INFORMASI ARTIKEL | ABSTRAK |
|---|---|
| **Kata Kunci:**<br>manajemen pendidikan, visi Indonesia 2045, mutu pendidikan | *Indonesia yang berdaulat, maju, adil, dan makmur telah ditetapkan sebagai Visi Indonesia 2045. Sebuah visi yang merangkum tujuan besar suatu negara dalam menempatkan dirinya baik dalam konteks internal maupun eksternal sebagai bagian dari peradaban dunia. Visi ini sekaligus menjadi acuan sekaligus dasar dalam pembangunan sumber daya manusia (SDM) untuk mendukung perwujudannya. Pembagian urusan pendidikan antara Pemerintah Pusat dan Pemerintah Daerah menjadi tantangan tersendiri dalam mewujudkan SDM yang bermutu. Pemerintah Pusat di satu sisi mempunyai kewenangan sebagai regulator standarisasi sistem pendidikan nasional. Pemerintah Daerah di sisi lain menghendaki sistem pendidikan yang mengakomodasi kebutuhan, keunikan, dan keunggulan daerahnya. Lalu bagaimana peran keduanya dalam penyediaan pendidikan yang bermutu? Apa saja tantangan yang harus dihadapi? Studi literatur ini akan mengupas dimensi kelayakan dalam kualitas pendidikan, serta peran Pemerintah Pusat dan Daerah dalam menyediakan pendidikan bermutu demi mewujudkan Visi Indonesia 2045.* |
| | ***ABSTRACT*** |
| **Keywords:**<br>*Education management, the vision of Indonesia 2045, education quality* | *A sovereign, advanced, just, and prosperous Indonesia has been designated as the Vision of Indonesia 2045. A vision that encapsulates the great goals of a country in positioning itself both in an internal and external context as part of world civilization. This vision is at the same time a reference as well as a basis in the development of human resources (HR) to support its realization. The division of educational affairs between the Central Government and Regional Governments is a challenge to creating quality human resources. The Central Government on the one hand has the authority as a regulator of the standardization of the national education system. Regional Governments on the other hand want an education system that accommodates the needs, uniqueness, and advantages of their region. Then what is the role of both in the provision of quality education? What are the challenges that must be faced? This literature study will explore the feasibility dimension in the quality of education, as well as the role of the Central and Regional Governments in providing quality education to realize the Vision of Indonesia 2045.* |
| ***Corresponding Author:***<br>E-mail:<br>ani_dikbud@bangkaselatankab.go.id<br><br>DOI: 10.5281/zenodo.7619138<br>(cc) BY-NC-ND | |

## A. Pendahuluan

Visi Indonesia 2045, yakni Indonesia yang berdaulat, maju, adil, dan makmur (Bappenas, 2020: 3). Visi tersebut ditunjang dengan pilar pendukung pencapaiannya yaitu: (1) membangun manusia dan menguasai iptek, (2) ekonomi berkelanjutan, (3) pembangunan yang merata, serta (4) ketahanan nasional dan tata kelola kepemerintahan yang mantap. Keempat pilar tersebut dibangun di atas Pancasila dan UUD 1945 sebagai dasar berbangsa, bernegara, dan berkonstitusi.

Salah satu pilar utama pendukung pencapaian visi adalah membangun manusia serta menguasai iptek. Secara lebih rinci, pilar ini bertujuan untuk mencapai peningkatan kualitas manusia Indonesia dengan pendidikan yang semakin tinggi dan merata; kebudayaan yang kuat; derajat kesehatan, usia harapan hidup, dan kualitas hidup yang semakin baik; produktivitas yang tinggi; serta kemampuan penguasaan ilmu pengetahuan dan teknologi yang luas (Bappenas, 2020: 8). Pembangunan manusia serta penguasaan iptek merupakan urusan wajib pendidikan sebagai amanat Undang-Undang.

Undang-Undang Nomor 23 Tahun 2014 tentang Pemerintahan Daerah telah secara jelas memberikan pembagian urusan antara Pemerintah Pusat dan Daerah. Pendidikan merupakan urusan wajib bidang pelayanan dasar. Dalam undang-undang tersebut dijelaskan bahwa pendidikan anak usia dini, pendidikan dasar, dan pendidikan non-formal merupakan kewenangan Pemerintah Kabupaten/Kota. Pendidikan menengah dan pendidikan khusus menjadi kewenangan Pemerintah Provinsi. Pendidikan tinggi dan penetapan standar nasional pendidikan menjadi kewenangan Pemerintah Pusat.

Urusan layanan pendidikan dasar yang menjadi kewenangan pemerintah kabupaten/kota meliputi: penetapan kurikulum muatan lokal pendidikan dasar, pemindahan pendidik dan tenaga kependidikan di dalam wilayah kabupaten/kota, penerbitan ijin pendidikan dasar yang diselenggarakan masyarakat, dan pembinaan bahasa dan sastra yang penuturnya dalam daerah kabupaten/kota (Arwildayanto, Lamatenggo, & Sumar, 2017: 93).

Pembagian urusan layanan pendidikan ini dimaksudkan agar masing-masing dapat fokus dalam memberikan pelayanan yang bermutu bagi masyarakat. Pendidikan merupakan salah satu urusan pemerintah daerah yang amat penting, menyangkut hajat hidup orang banyak, menentukan masa depan anak bangsa, dan juga akan ikut menentukan maju-mundurnya daerah itu sendiri dalam jangka panjang, jika diukur dari kualitas sumber daya manusia yang dimiliki daerah tersebut sebagai hasil dari proses pendidikan yang diurusnya.

Urusan pendidikan dasar seharusnya lebih mempunyai otonomi dalam pelaksanaannya karena kewenangannya berada di kabupaten/kota. Kepala daerah memiliki kewenangan yang penuh dalam menentukan kualitas sekolah di daerahnya masing-masing melalui perekrutan kepala sekolah, guru, dan siswa, pembinaan profesionalisme guru, penentuan sistem evaluasi, dan sebagainya (Hasibuan, 2017: 90). Akan tetapi dalam kenyataannya pemerintah kabupaten/kota masih belum mampu membawa pendidikan di daerahnya menjadi lebih maju dan bermutu (Hartono, 2015; Nasution, 2010; Ridwan & Sumirat, 2021).

Beberapa kendala yang menyebabkan belum majunya pendidikan di daerah antara lain kurangnya visi kepala daerah untuk memajukan pendidikan (Harahap, 2016), sumber daya daerah yang belum memadai baik SDM maupun SDA (Maisyanah, 2018), masih tingginya campur tangan Pemerintah Pusat terhadap pelaksanaan otonomi pendidikan di daerah (Hartono, 2015; Ridwan & Sumirat, 2021), serta rendahnya keterlibatan masyarakat dalam memajukan pendidikan daerah (Armansyah, 2016).

Kondisi pendidikan dasar di daerah sangat beragam, keterlayanan dan kualitasnya. Ada daerah yang telah menuju kualitas sebagai prioritas, namun ada juga daerah yang masih berkutat dengan pemerataan perolehan kesempatan pendidikan. Jika kondisi ini terus dibiarkan tanpa penyelesaian maka jelas akan merugikan kepentingan publik karena pendidikan menyangkut pemenuhan hajat hidup orang banyak. Oleh karena itu perlu segera ditemukan solusi yang ideal untuk dapat mengatasi berbagai kendala yang menyebabkan peningkatan mutu pendidikan dasar tidak beranjak naik.

## B. Metode Penelitian

Riset ini mengambil bentuk umum dari studi literatur yang berusaha mengumpulkan landasan peraturan perundang-undangan sebagai patokan dalam mengidentifikasi bentuk ideal dari pelayanan pendidikan. Selanjutnya mencari hasil riset-riset terdahulu untuk mengumpulkan kendala dan usaha dalam memajukan kualitas pendidikan khususnya di daerah. Ketiganya kemudian dipadukan untuk memberikan solusi dalam upaya mewujudkan visi Indonesia 2045 melalui pendidikan yang bermutu.

## C. Hasil

Pemerintah berkewajiban menjamin rakyatnya untuk mendapatkan pendidikan yang layak sesuai dengan amanat Undang-Undang Dasar 1945. Diksi 'layak' dalam penerapannya mempunyai tingkatan keterlaksanaan. Layak dalam kesetaraan memperoleh pendidikan, layak mendapatkan fasilitas penunjang pendidikan, dan layak untuk mendapatkan layanan pendidikan yang berkualitas.

1. Layak dalam Kesetaraan Memperoleh Pendidikan

Layak dalam kesetaraan memperoleh kesempatan pendidikan dapat diartikan dengan pemerataan kesempatan mendapatkan pendidikan oleh seluruh warga negara tanpa terkecuali, tanpa memandang status sosial, ekonomi, politik, gender, dan budaya. Segala bentuk diskriminasi tidak diperkenankan dan melanggar hukum. Tidak ada satu wargapun yang tidak mendapatkan layanan pendidikan bagaimanapun keadaannya. Pada konteks ini pendidikan dapat bermakna luas tidak hanya sebatas pendidikan formal namun juga pendidikan non-formal. Keterampilan dasar yang harus diajarkan setidaknya mencakup membaca, berhitung, dan menulis (Shields, Liam, Newman, & Satz, 2017: 17). Pada tahap ini fasilitas penunjang dan kualitas layanan pendidikan tidak begitu dipersoalkan karena keterbatasan anggaran yang dimiliki suatu daerah dihadapkan pada kondisi geografis yang ekstrim sehingga menyebabkan pemerintah daerah kesulitan untuk membangun sarana pendidikan yang memadai.

Pemerintah daerah tidak boleh mengabaikan hak-hak mendapatkan pendidikan bagi warganya bagaimanapun keterbatasan anggaran yang dimilikinya. Selain itu di tengah keterbatasannya, pemerintah daerah juga dituntut untuk mendukung keberhasilan program nasional wajib belajar 9 tahun. Warga yang wajib dilayani tidak terbatas pada warga yang berada dalam usia sekolah, namun juga terhadap seluruh warga dengan berbagai usianya yang belum menuntaskan pendidikan dasar 9 tahun. Pada kenyataannya pada tahap inipun masih banyak menyisakan permasalahan yang ditandai dengan masih rendahnya rerata lama sekolah yang ditempuh.

2. Layak Mendapatkan Fasilitas Penunjang Pendidikan

Layak mendapatkan fasilitas pendidikan dapat diartikan bahwa warga negara telah menikmati pendidikan sesuai dengan standar nasional pendidikan (SNP) yang ditetapkan. Fasilitas penunjang pendidikan tidak hanya bermanfaat bagi siswa dalam membantu belajarnya, melainkan juga bagi pendidik untuk menunjang kelancaran pelaksanaan tugasnya (Kristiawan, Safitri, & Lestari, 2017: 62). Tahap ini sedikit lebih menguntungkan bagi warga, karena setidaknya mereka dapat merasakan sedikit "kemewahan" bersekolah. Meskipun dengan terpenuhinya SNP hanya berarti memenuhi standar minimal pelayanan yang ditetapkan, namun telah berorientasi pada ketercapaian mutu pada tahap awal.

Sepintas hal ini telah memenuhi rasa keadilan, namun ternyata masih menyisakan permasalahan lain, misalnya bagaimana dengan siswa-siswa yang mempunyai kecerdasan/bakat istimewa, apakah mereka dapat terlayani dengan baik? Atau pertanyaan tentang bagaimana sekolah menyiapkan keunggulan lulusannya?

3. Layak Mendapatkan Layanan Pendidikan yang Berkualitas

Mendapatkan layanan pendidikan yang berkualitas berarti terlampauinya standar nasional pendidikan. Pada tahap ini dapat berharap pada ketercapaian kualitas hasil pendidikan yang dilaksanakan. Namun apakah sebagian besar warga negara telah mendapatkan pendidikan dasar yang berkualitas? Data pencapaian kualitas pendidikan Indonesia dapat dilihat dengan membandingkan hasil pendidikan dengan negara lain pada satu atau beberapa aspek sebagai hasil pelaksanaan pendidikan seperti PIRLS, PISA atau TIMMS. Jenis studi penelitian internasional ini dilakukan setiap beberapa tahun sekali untuk mengidentifikasi tingkat pencapaian pendidikan di negara-negara yang berpartisipasi (Alfirevic, Burusic, Pavicic, & Relja, 2016: 19). Hasilnya selalu saja menempatkan Indonesia sebagai negara dengan peringkat nyaris terendah. Jadi apa yang menyebabkan kualitas pendidikan di Indonesia masih rendah meskipun layanannya sudah melebihi standar nasional pendidikan?

4. Penyebab Rendahnya Mutu Pendidikan Dasar

a. Kurikulum Nasional yang Tidak Memperhatikan Kondisi Daerah

Pemerintah pusat berwenang menetapkan kurikulum nasional yang harus dijadikan pedoman pengembangan kurikulum pendidikan dasar. Dalam konteks otonomi daerah kurikulum pendidikan tidak dapat disamaratakan dengan daerah-daerah lain yang kondisinya berbeda. Hal ini akan mempengaruhi ketercapaian output hasil pendidikan yang di dalamnya kompleks. Kurikulum akan melahirkan suatu tuntutan atas terselenggaranya suatu rencana pembelajaran yang keberhasilannya tergantung seberapa kuat dukungan dari sumber daya pendukung yang ada. Belum lagi kaitannya dengan relevansi kurikulum terhadap kenyataan bahwa masyarakat Indonesia sangat heterogen (Ridwan & Sumirat, 2021: 102). Heterogenitas ini dapat dilihat dari keanekaragaman budaya, adat, suku, sumber daya alam dan bahkan sumber daya manusianya. Setiap daerah mempunyai kesiapan dan kemampuan yang berbeda dalam pelaksanaan kurikulum nasional, desentralisasi pendidikan.

b. Kurang Apiknya Penataan Sumber Daya Manusia Pendidikan

Sudah menjadi hal yang jamak bahwa kepala daerah mempunyai kewenangan yang otonom dalam menempatkan SDM di bawah kepemimpinannya. *Like and Dislike*, KKN, dan perilaku menyimpang dalam penempatan SDM pendidikan sangat mempengaruhi kinerja lembaga pendidikan. Sebagaimana yang disampaikan oleh Maisyanah bahwa ketika pejabat yang berwenang mengambil kebijakan dalam bidang pendidikan tidak berkualitas dan tidak memahami pendidikan secara holistik, maka keputusan yang diambil cenderung tidak berdampak positif terhadap perkembangan dunia pendidikan (Maisyanah, 2018: 7).

Bagaimanapun penempatan sumber daya manusia tidak sesuai dengan latar belakang pendidikan dan profesionalnya dan banyak tenaga kependidikan yang latar belakang pendidikannya tidak relevan kerja yang ditekuninya merupakan penghambat dalam pelaksanaan pendidikan di daerah (Ridwan & Sumirat, 2021: 103). Sebagai contoh bagaimanapun dekatnya hubungan dengan kepala daerah, seorang pegawai dinas pemakaman tentu tidak cocok untuk diangkat menjadi kepala dinas pendidikan.

c. Lemahnya Sinkronisasi dan Koordinasi Kebijakan Pendidikan antara Pemerintah Pusat dan Daerah

Kebijakan pendidikan yang diterapkan di Indonesia pada umumnya masih berjalan sendiri-sendiri, tidak terjadi sinkronisasi dan koordinasi yang jelas. Hal ini dapat disebabkan oleh besarnya otonomi dan jelasnya pemisahan tugas dan kewenangan yang menyertainya. Kepala dinas pendidikan kabupaten/kota tentu akan lebih menuruti kebijakan yang dipilih oleh kepala daerahnya daripada menaati aturan yang dikeluarkan seorang menteri.

Sebagai contoh menteri mengeluarkan kebijakan peniadaan ujian sekolah berstandar nasional, namun ada pemerintah daerah yang masih tetap melaksanakannya meskipun dalam bentuk yang lain (medcom.id, 2021). Atau ada kepala daerah yang menolak SKB 3 Menteri tentang seragam sekolah (kompas.com, 2021), bahkan sampai MA membatalkan SKB 3 Menteri tersebut (cnnindonesia.com, 2021). Beberapa contoh yang telah dikemukakan tentu semakin memperlihatkan kejanggalan dan ketidakharmonisan tata pemerintahan di Indonesia.

d. Tingginya Penyelewengan Anggaran dan Penyediaan Fasilitas Pendidikan

Tingginya tingkat penyelewengan anggaran yang dilakukan di tingkat kabupaten/kota (Utami, 2018:40) menyebabkan tidak maksimalnya penganggaran dan penyediaan fasilitas pendidikan di daerah. Padahal banyak opini dibangun untuk menyakinkan publik bahwa kepala daerah sudah memenuhi kewajibannya dalam menyediakan anggaran pendidikan sebesar 20% dari keuangan daerah (Ridwan & Sumirat, 2021: 104). Padahal sesuai amanat undang-undang, besaran anggaran 20% itu minimal. Jika menempatkan sektor pendidikan sebagai sektor unggulan dalam pembangunan daerah tentu akan mempertimbangkan alokasi anggaran pendidikan yang lebih besar dari 20%. Akibatnya program-program peningkatan kualitas pendidikan jarang tersentuh.

e. Esensi Pembelajaran Terkikis dalam Pelaksanaannya

Para guru masih banyak yang apatis, statis dalam menanggapi pembaharuan atau perubahan pendidikan, mereka masih banyak terbelenggu pada sistem pembelajaran konvensional yang lebih menekankan pada pemberian informasi serta dan tidak memperhatikan aspek afektif dan psikomotorik (Ridwan & Sumirat, 2021: 105). Kebijakan pendidikan nasional yang terpusat dan cenderung menyeragamkan, secara jelas bertentangan dengan realita masyarakat Indonesia yang majemuk di berbagai daerah. Sangat beralasan jika pada akhirnya proses pembelajaran yang terjadi hanya berfokus pada aspek kognitif yang menekankan formalisme dan pada saat yang sama, cenderung mengabaikan ranah afektif dan psikomotorik.

f. Kurangnya Dukungan Masyarakat dalam Kemajuan Pendidikan Daerah

Komite sekolah sebagai perwujudan dukungan masyarakat terhadap pendidikan belum berfungsi maksimal. Selama ini komite sekolah hanya dianggap sebagai alat kelengkapan semata tanpa memberikan bantuan yang bersifat signifikan terhadap kebutuhan sekolah. Problem yang sama, yaitu komite, atau sebagian anggota komite belum mengetahui secara utuh tentang hakikat pendidikan itu sendiri (Maisyanah, 2018: 11). Permasalahan ini menyebabkan pendidikan tanpa dukungan sekaligus pengawasan yang benar dalam pelaksanaannya, sehingga tidak terjadi kontrol sosial yang benar atas penyelenggaraan pendidikan. Masyarakat hanya menerima saja apa yang dilakukan lembaga pendidikan tanpa berusaha meluruskan atau membenahi jika ada upaya penyelewengan.

## D. Pembahasan

Bangsa Indonesia tengah diliputi euforia dalam menyambut era kebangkitan nasional didasarkan pada berbagai peluang yang mendekat. Keunggulan di bidang sumber daya alam dan energi, tren pertumbuhan ekonomi, prospek poros perdagangan dunia, pengaplikasian tren perkembangan teknologi, serta bonus demografi digadang-gadang akan membawa Indonesia ke puncak peradaban dan keunggulan dalam tataran persaingan global.

Optimisme tersebut tergambar dalam visi Indonesia 2045, yakni Indonesia yang berdaulat, maju, adil, dan makmur (Bappenas, 2020: 3). Pilar pendukung pencapaian visi tersebut yaitu: (1) membangun manusia dan menguasai iptek, (2) ekonomi berkelanjutan, (3) pembangunan yang merata, serta (4) ketahanan nasional dan tata kelola kepemerintahan yang mantap. Keempat pilar tersebut dibangun di atas Pancasila dan UUD 1945 sebagai dasar berbangsa, bernegara, dan berkonstitusi. Salah satu pilar utama pendukung pencapaian visi adalah pembangunan manusia serta penguasaan iptek. Secara lebih rinci, pilar ini bertujuan untuk mencapai peningkatan kualitas manusia Indonesia dengan pendidikan yang semakin tinggi dan merata; kebudayaan yang kuat; derajat kesehatan, usia harapan hidup, dan kualitas hidup yang semakin baik; produktivitas yang tinggi; serta kemampuan penguasaan ilmu pengetahuan dan teknologi yang luas (Bappenas, 2020: 8).

Strategi yang dapat dilakukan Pemerintah (Pusat dan Daerah) dalam upaya peningkatan mutu pendidikan dasar, sebagai berikut:

1. Mengurangi kebijakan sentralistik dalam urusan pendidikan dasar

Urusan pendidikan dasar merupakan kewenangan Pemerintah Kabupaten/Kota. Namun demikian tidak berarti bahwa Pemerintah Pusat tidak ikut ambil bagian di dalamnya. Peran Pemerintah Pusat di antaranya adalah menentukan standar nasional pendidikan dan kebijakan sebagai acuan umum terhadap langkah pelaksanaan pendidikan di daerah. Adapun dalam penerapannya harus memberikan ruang bagi Pemerintah Kabupaten/Kota untuk berimprovisasi, berinovasi, dan berkreasi untuk mewujudkan pendidikan dasar yang bermutu di daerahnya.

2. Kebijakan pendidikan harus senda seirama antara Pemerintah Pusat dan Pemerintah Daerah

Otonomi berarti upaya pemenuhan hak dan kewajiban yang dilakukan melalui desentralisasi atau proses untuk mencapai hak dan kewajiban tersebut. Asas keadilan harus diutamakan untuk mencapai tujuan yang diharapkan. Pemerintah Pusat tidak boleh memaksakan kehendaknya kepada Pemerintah Daerah. Sebaliknya Pemerintah Daerah harus konsisten melaksanakan aturan yang telah dibahas dan disepakati bersama antara Pemerintah Pusat dan Pemerintah Daerah.

3. Pelibatan seluruh lapisan masyarakat dalam upaya peningkatan mutu pendidikan dasar

Perlu pelibatan seluruh lapisan masyarakat dalam upaya peningkatan mutu pendidikan dasar (Maisyanah, 2018: 9). Kepercayaan pemerintah pusat yang masih sangat kurang terhadap kredibilitas pemerintah daerah dalam menata sistem pendidikannya memperlebar masalah relevansi pendidikan. Keseimbangan input, proses, dan hasil pendidikan yang senada dengan kondisi objektif setiap daerah mutlak menjadi perhatian dalam menentukan kebijakan pendidikan nasional.

Kunci yang mampu mewujudkan tujuan pendidikan di indonesia bukan hanya menjadi tanggung jawab pemerintah saja, melainkan juga tanggung jawab dan kerjasama antara pemerintah, masyarakat dan elemen terkecil dan terpenting dalam mewujudkan pendidikan di Indonesia ini adalah keluarga. Semuanya harus mampu bertindak sebagai praktisi sekaligus mampu mengontrol secara bersama-sama dalam upaya meningkatkan kualitas pendidikan dan mencapai tujuan pendidikan baik secara mikro maupun secara makro.

Studi literatur ini membuahkan rekomendasi yang dapat dilakukan Pemerintah Pusat dan Daerah dalam meningkatkan mutu pendidikan dasar:

1. Sinkronisasi kebijakan antara Pemerintah Pusat dan Daerah dalam bidang pendidikan merupakan kegiatan yang wajib dilakukan. Pemerintah pusat harus melibatkan pemerintah daerah sebelum memutuskan suatu kebijakan agar benar-benar diketahui pembagian tugasnya secara jelas, tidak hanya sebuah gebrakan yang hangat-hangat tahi ayam.

2. Pemerintah harus berkaca pada sejarah dan tidak membuat jalannya sendiri hanya untuk memastikan pada masa jabatannya ada suatu ciri khas kebijakan yang mencerminkan dirinya, namun mengabaikan keberhasilan secara keseluruhan.

3. Pemerintah pusat berwenang untuk menetapkan standar dan acuan yang jelas sehingga pemerintah kabupaten/kota mampu mengembangkannya menjadi kebijakan yang lebih terintegrasi dengan perencanaan nasional dan daerah dalam upaya untuk mencapai tujuan pendidikan nasional.

4. Pembangunan pendidikan harus selalu berorientasi pada kepentingan masyarakat. Masyarakat di daerah harus diposisikan sebagai subjek sekaligus pihak yang mendapatkan manfaat dari program dan kegiatan yang akan dilaksanakan. Sehingga program pembangunan diarahkan untuk kegiatan yang bertujuan memenuhi kebutuhan praktis dan strategis masyarakat yang hasil dan dampaknya dapat dirasakan langsung oleh masyarakat kelas bawah hingga masyarakat puncak.

5. Program pendidikan harus sesuai dengan kebutuhan masyarakat. Proses perencanaan, pelaksanaan sampai kepada pengawasan melibatkan masyarakat. Seluruh aspirasi, kebutuhan daerah dan masyarakat harus terakomodir dan hasil-hasil pelaksanaan pendidikan dapat dinikmati secara langsung serta dapat memberdayakan masyarakat di daerah.

6. Kurikulum seyogyanya memberikan ruang bagi Pemerintah Daerah untuk berimprovisasi sesuai adat dan budaya masyarakat setempat yang terpelihara dan berkembang dalam masyarakat sebagai sebuah kerifan lokal yang memperkaya khasanah budaya bangsa dalam kerangka orientasi lokal, nasional, regional, dan global. Ini akan menjadi keunggulan lokal sebagai hasil dari peningkatan mutu pendidikan dasar.

## E. Penutup

Visi Indonesia 2045 tinggal di depan mata. Perlu upaya yang serius untuk dapat mencapainya. Pendidikan yang bermutu sebagai salah satu kunci menyukseskan Visi Indonesia 2045 tidak dapat ditawar lagi. Perlu strategi yang komprehensif, berkesinambungan, dan berbudaya dari berbagai pihak terutama Pemerintah sebagai regulator yang akan menentukan langkah bagi para penyelenggara pendidikan.

## F. Ucapan Terima Kasih

Penulis mengucapkan terima kasih kepada para pihak yang telah membantu penelitian ini, terutama kepada Kepala Dinas Pendidikan dan Kebudayaan Kabupaten Bangka Selatan atas ijin yang diberikan.